\documentclass[11pt,
]{article}
\usepackage[]{graphicx}
\usepackage{makeidx}
\newcommand{\mathsym}[1]{{}}

\newcommand{\bi}{\bibitem}
\newcommand{\be}{\begin{eqnarray}}
\newcommand{\ee}{\end{eqnarray}}
\newcommand{\rar}{\rightarrow}

\usepackage[]{caption}
\def\gsim{\mathrel{\raise.3ex\hbox{$>$\kern-.75em\lower1ex\hbox{$\sim$}}}}
\def\lsim{\mathrel{\raise.3ex\hbox{$<$\kern-.75em\lower1ex\hbox{$\sim$}}}}

\begin{document}

\begin{center}
{\bf \large{
BARYOGENESIS AND COSMOLOGICAL ANTIMATTER.
}} \\ \vspace{0.5cm}
{ Alexander D. Dolgov \\
\it{Istituto Nazionale di Fisica Nucleare, Sezione di Ferrara, 
I-44100 Ferrara, Italy \\
Dipartimento di Fisica, Universit\`a degli Studi di Ferrara, 
I-44100 Ferrara, Italy \\
Institute of Theoretical and Experimental Physics, 113259 Moscow, 
Russia }}

\begin{abstract}
Possible mechanisms of baryogenesis are reviewed. Special attention is payed 
to those which allow for creation of astronomically significant domains or
objects consisting of antimatter. Observational manifestations of cosmological
antimatter are discussed.
\end{abstract}
\end{center}

\section{Introduction}

Theoretical prediction of antimatter made by Paul Dirac 80 years ago is one of the
most impressive discoveries of quantum field theory~\cite{dirac-28}. 
A  symmetry between matter and antimatter led Dirac to a 
natural suggestion that ``maybe there exists a completely new universe
made of antimatter''. Surprisingly a very similar statement was done 30 years 
before Dirac by another English physicist Arthur Schuster who made a brilliant
but at that time wildly speculative guess that
there might be entire solar systems, made of antimatter and indistinguishable 
from our Solar system (probably by light) but capable to annihilate and produce 
enormous energy. To avoid abundant antimatter around us, Schuster assumed that 
world and antiworld were gravitationally repulsive.

Our present point of view on existence of such astronomically large systems
is very much different, even opposite. 
Now we know for sure that antimatter exists but believe that there are very 
few antiparticles in the universe, except for antineutrinos, which, together with 
neutrinos, are the most abundant massive particles in the universe. As for ``real''
antimatter, i.e. antiprotons or positrons, the dominant point of view is 
that they are
too rare to make any macroscopic objects. 

However it is not excluded that Dirac and Schuster were right and there are
whole antimatter worlds, or solar-like systems, or just macroscopically 
large antimatter objects. It is discussed below if there may exist 
whole galaxies made of antimatter or could galaxies or the Galaxy consist 
predominantly of matter with large clumps of antimatter which are potentially observable.
It is argued that both natural theory and existing observations allow for that.

These lectures are based on the works with J. Silk~\cite{ad-js},
C. Bambi~\cite{cb-ad}, M. Kawasaki and N. Kevlishvili~\cite{ad-mk-nk}, see 
also~\cite{ad-thuile}. Recent theoretical activity was stimulated by the 
existing: Pamela, BESS, AMS, and future AMS-02 (2009), PEBS (2010),
CAPS (2013) programs for search of cosmic antimatter, according to the
talk by P. Picozza at TAUP 2007~\cite{picozza}.

The content of the lectures is the following. In the next section the basic
ideas of baryogenesis are formulated and their ability to give birth to 
astronomically significant amount of antimatter is discussed. Different 
mechanisms of CP-violation which are favorable for antimatter creation are
enumerated in sec.~\ref{s-CP}. Observational data are briefly described in
sec.~\ref{s-data}. A specific mechanism of creation of compact antimatter objects, 
which nevertheless may have a larger total mass than the observed baryons, 
is presented in sec.~\ref{s-anti-creation}. Cosmological evolution of such
(anti)baryonic bubbles is considered in sec.~\ref{s-inhom}. Observational 
manifestations of abundant antimatter in the Galaxy is described in
sec~\ref{s-antigal}. At last, in sec.~\ref{s-concl} we conclude.

\section{Cosmological baryogenesis \label{s-bs}}

It is strongly believed that the universe is
populated by matter and that 
the observed antimatter is of the secondary origin.
Observations and the simplest scenarios of baryogenesis 
do not contradict this assertion.  
The standard mechanism of baryogenesis is based on the following three
principles as it was formulated in 1967 by 
A.D. Sakharov~\cite{sakharov}:\\
1.  Non-conservation of baryons. It is predicted theoretically
by grand unification~\cite{gut-B} and even by the standard electroweak 
theory~\cite{ew-B}.
Moreover, in a sense it is confirmed ``experimentally''  by cosmological
inflation because the latter is impossible with the conserved
baryonic density equal to the measured one. It is interesting how the 
same observational fact of our existence led to opposite conclusions:
30 years ago it was: ``we exist, thus baryons are conserved.'',
while now it is: ``we exist, thus baryons are NOT conserved.''.\\
2. Breaking of symmetry between particles and antiparticles, 
i.e. of C and CP. CP-violation was observed in experiment in 
1964~\cite{cp-break}. Breaking of C-invariance was found earlier
immediately after discovery of parity non-conservation~\cite{parity}. \\
3. Deviation from thermal equilibrium. This is fulfilled in nonstationary,
expanding universe for massive particles or due to possible first order 
phase transitions. 

Neither of the three conditions is obligatory, 
and successful baryogenesis could be realized without any of them~\cite{ad-bs}, 
but the corresponding scenarios are somewhat more complicated.

The list of the available baryogenesis scenarios is surely not a short
one. There is plethora of models each 
being able to explain the single observed number:
\be 
\beta_{observed} = \frac{N_B-N_{\bar B}}{N_\gamma} 
\approx 6\times 10^{-10}
\label{beta-obs}
\ee
The standard prediction of different models is $ \beta = const$, 
which makes impossible to distinguish between the models.
Much more interesting are the models withspatially varying 
asymmetry, $ {\beta = \beta (x)}$ and especially with negative
$\beta (x)$ in some astronomically large regions, i.e. with
significant amount of antimatter, maybe not too far from us.
The scenarios of baryogenesis include:

I. {\it Baryogenesis by heavy particle decays.} This was 
historically first model~\cite{sakharov,kuzmin}; the
references to subsequent works can be found e.g. in reviews~\cite{ad-bs,ad-yz}.
Due to C and CP non-conservation the {\it partial} widths of decays into 
channels with different baryonic numbers should be different. For example:
\be 
\Gamma(X \rar q \bar l) \neq \Gamma(\bar X \rar \bar q l),
\label{Gamma-X-ql}
\ee
where $X$ is a heavy particle and $q$ and $l$ are correspondingly
a quark and a lepton. Such a mechanism can be realized in particular in
grand unification models with $X$ being a gauge boson of grand unification
with $m_X \sim 10^{16}$ GeV.

Deviation from thermal equilibrium can be easily estimated from the 
kinetic equation in expanding universe and is equal to:
\be 
\frac{\delta f}{f_{eq}} = \frac{H m^2_X }{\Gamma T E}
\approx \frac{10 m_X}{\alpha m_{Pl}},
\label{delta-f}
\ee
where $H$ is the Hubble parameter and $g_* \sim 100$ is the number of
particle species in the primeval plasma. The baryogenesis is most 
efficient at $T\sim m_X$ (in fact about an order of magnitude smaller)
and an optimistic estimate of the resulting asymmetry is
\be 
\beta \approx \frac{\Delta \Gamma}{\Gamma}\,\frac{\delta f}{f_{eq}}\,
\frac{ n_X}{n_0} \sim \frac{10 m_X}{m_{Pl}}\,\frac{n_X}{n_0},
\label{beta-gut}
\ee
where $n_0\sim 0.1 T^3$ is the number density of massless particles 
and $m_{Pl}$ is the Planck mass.

II. {\it Electroweak baryogenesis~\cite{krs}.} 
The standard $ SU(2)\times U(1)$ model
possesses all the necessary ingredients for baryogenesis: 
C and CP violation, non-conservation of baryons
(through the chiral anomaly)~\cite{ew-B},
and significant deviation from thermal equilibrium, if the phase transition
from the electroweak symmetric phase to the phase with broken symmetry is
1st order. However, heavy Higgs boson, $m_H > 100$ GeV leads to the second order
phase transition and to very weak deviation from thermal equilibrium. Moreover,  
CP violation in the minimal standard model is by far too weak to 
ensure efficient baryogenesis, see e.g. discussion in ref.~\cite{ad-varenna}

A possible cure may be TeV scale gravity~\cite{TeV-gravity}, with $m_{Pl}\sim $ TeV,
which allows both significant deviation from thermal equilibrium, even
without 1st order phase transition, see eqs. (\ref{delta-f},\ref{beta-gut}),
and a stronger CP-violation~\cite{TeV-grav-EW}. 

III. {\it Baryo-through-lepto-genesis~\cite{fuku-yana}}. 
This is a combination of the 
scenarios I and II. First, at temperatures about $10^{10}$ GeV lepton asymmetry
was generated in the decays of a heavy Majorana fermion and subsequently the
lepton asymmetry was redistributed between baryon and lepton ones approximately
in equal share by the equilibrium electroweak processes which conserve $(B-L)$
but break $(B+L)$. For a recent review see e.g. ref.~\cite{lgns}.

IV. {\it  SUSY condensate (Affleck-Dine) baryogenesis~\cite{affleck}.} 
In supersymmetric
models there must exist scalar superpartners of baryons or leptons $\chi$. The 
potentials, $U(\chi)$,
of such scalar fields have some flat directions along which the fields
can develop a non-zero vacuum expectation values e.g. due to quantum
fluctuations during inflation. After inflation is over and non-zero mass of
$\chi$ is generated, $\chi$ evolves down to the mechanical equilibrium point
$\chi = 0$. If the potential, $U(\chi)$ is not  symmetric with respect to the 
phase rotation, $\chi \rightarrow \exp(i\Theta \chi)$, i.e. baryonic charge is
not conserved in $\chi$ self-interaction, then in the process of relaxation 
of $\chi$ down to zero, the field starts to ``rotate'' around the origin, i.e.
it acquires non-vanishing and typically large baryonic charge.  
Subsequent $B$-conserving decay of 
$\chi$ into quarks and/or leptons transform baryon (or lepton) asymmetry 
in the $\chi$ sector into that in the quark sector.
For more detail see below Sec.~\ref{s-anti-creation}.
In contrast to other scenarios of baryogenesis, this one normally leads
to quite high value of $ {\beta = n_B/n_\gamma\sim 1}$ and theoretical efforts 
are needed to diminish it down to the observed value.
This mechanism is especially favorable for creation of astronomically large
antimatter domains.

V. {\it Spontaneous baryogenesis~\cite{spont-BS}.}
It is assumed that theory is symmetric with respect to 
spontaneously broken global ${U(1)}$ symmetry associated with ${B}$.
In the broken phase there appears a massless or light
Goldstone field $ \theta$. Its Lagrangian has the form:
\be 
{\cal L} = \eta^2 (\partial \theta )^2 + {\partial_\mu \theta j^B_\mu} 
- V(\theta) + 
i\bar Q \gamma_\mu \partial_\mu Q + i\bar L \gamma_\mu \partial_\mu L +
(g \eta \bar Q L + h.c.),
\label{L-theta}
\ee
The time derivative of $\theta$ looks as a 
chemical potential but strictly speaking this is not so because chemical
potential is introduced into Hamiltonian and such an addition to the Hamiltonian
coincides with that to the Lagrangian only if it does not contain
derivatives~\cite{ad-kf} which is surely not true in this case. Nevertheless 
the model may quite efficient for creation of cosmological baryon asymmetry.
It is interesting that baryogenesis is possible in thermal equilibrium. 
This scenario is also favorable for creation of cosmologically significant
amount of antimatter. However, as far as I know, all scenarios of antimatter 
creation based on spontaneous baryogenesis suffer from too
large isocurvature density perturbations at large scales which are
forbidden by CMBR.

VI. {\it Baryogenesis through evaporation of primordial black 
holes~\cite{zeld-bh}.} Though the Hawking process creates thermal equilibrium
spectrum of the emitted particles at the black hole (BH) horizon, the particle
propagation in the gravitational field of the BH distorts it. It makes possible
to create an excess of particles over antiparticles. A concrete model of such
generation of baryon asymmetry may be the following. Let BH emits a heavy X-boson
which decays in the vicinity of the BH horizon
into a light baryon and a heavy antibaryon and vise versa with different 
decay rates:
\be 
\Gamma(X \rar L + \bar H) \neq \Gamma( X\rar \bar L + H)
\label{gamma-L-H}
\ee
The probability of gravitational back-capture by the BH of $H$ is larger 
than  $L$ and it would lead to an excess of baryons in external space.

VII. {\it Space separation of $ B$ and $\bar B $.} 
The idea that baryons and antibaryons
are separated at cosmological scales due to mutual repulsion was suggested by 
Omn{\'e}s in 1970~\cite{omnes}. However, no repulsive interaction between baryons
and antibaryons of sufficient strength is known and the mechanism seems to be
excluded. A possibility of spatial separation of matter and antimatter
was reconsidered in multidimensional cosmologies 
with matter and antimatter living on different branes which might be quite
close to each other along the fourth dimension~\cite{dgp}.
Another suggestion which does not demand new physics was proposed in 
ref.~\cite{ez}, according to which quarks and antiquarks at QCD phase transition
can form astronomically small but macroscopic bubbles with large baryonic or
antibaryonic numbers. Due to CP violation a small misbalance between quark and
antiquark nuggets was created which is observed as the usual baryonic matter. Such
(anti)quark nuggets may form cosmological dark matter.
All these scenarios allow for baryonic charge conservation and as a result 
give rise to globally baryo-symmetric universe.

\section{CP violation in cosmology  \label{s-CP}}

Creation of astronomically significant antimatter depends not only of the
mechanism of baryogenesis but also on the mechanism of CP violation in
cosmology. For a review of cosmological CP-violation see ref.~\cite{ad-varenna}.
There are several distinct possibilities of breaking symmetry between
particles and antiparticles: \\
1.{\it  The usual explicit, by complex constants in Lagrangian.} Is is assumed
in majority of baryogenesis scenarios. In this case the baryon asymmetry is 
normally constant over space. However, under certain conditions 
the sign of the baryon excess 
is not prearranged by the sign of CP-symmetry breaking but is determined by the kinetics of
the processes~\cite{yokoyama}, and by intial conditions~\cite{grigoriev}. \\
2. {\it Spontaneous, induced by a non-zero vacuum expectation value of a complex
scalar field}~\cite{cp-spont}. Such a mechanism evidently 
leads to charge symmetric universe with mixed domains of matter and
antimatter~\cite{stecker}. The original scenario, however, suffers from very small
size of the domains.
Observations either exclude it or demand the nearest antimatter domain to be
beyond the present day horizon because of too high gamma ray background~\cite{CdRG} 
and also due to the domain wall problem~\cite{koz}. The first part of the problem can 
be resolved if the domains expanded exponentially after their formation~\cite{sato}
but the domain walls must be outside cosmological horizon.
\\
3.{\it Dynamical or stochastic CP breaking, by a complex scalar field which was 
displaced out of mechanical equilibrium point during BS and relaxed down to 
equilibrium now}~\cite{ad-bs,ad-varenna}. Evidently the domain wall problem in this
case is absent and the universe is not necessarily baryon-symmetric, though
significant antimatter is allowed. \\
4.{\it  A mixture of spontaneous and explicit CP-violation}~\cite{spont-explicit}.
It makes charge asymmetric universe with an arbitrary fraction of antimatter.
\\
5.{\it A mixture of all above} - everything which is not forbidden is allowed.

\section{Observations \label{s-data}}

Up to now no astronomically significant objects consisting antimatter 
have been detected. A little antiprotons and 
positrons in cosmic rays are most probably of 
the secondary origin. Quite suspicious is 
the observed intensive positron annihilation 0.511 MeV line from 
the galactic bulge~\cite{bulge} and possibly from the halo~\cite{halo}.
The abundant positrons may indicate to an existence of antimatter objects
(see below),
but most probably the annihilation line has less striking explanation.

As we have already mentioned, in charge symmetric universe 
the nearest antimatter domain should be 
practically at the present day horizon, {${ l_B >}$ Gpc}. 
An efficient annihilation on the domain boundaries at an early stage, 
due to positive feedback, would create too intensive cosmic 
gamma ray background~\cite{CdRG}.

{No significant amount of antimatter is observed in the Galaxy.}
Observed colliding galaxies  
or galaxies in the common cloud of intergalactic gas 
are dominated by the same kind of matter (or antimatter?). The absence of the
noticeable 100 MeV gamma radiation allows to limit the 
fraction of antimatter in the Bullet Cluster 
by $n_{\bar B}/n_B < 3\times 10^{-6}$~\cite{steigman-08}.
Analogously the nearest galaxy dominated by antimatter could not be closer 
than at $\sim$10 Mpc~\cite{steigman-76}. However we cannot say much about
galaxies outside of our supercluster.

However, one should keep in mind that these bounds are true if antimatter makes  
exactly the same type objects as the {\it observed} matter.
For example, compact objects made of antimatter may escape 
observations and be quite abundant and almost at hand.

\section{Anti-creation mechanism \label{s-anti-creation}}

A mechanism which leads to an abundant antimatter objects in the universe
and, in particular in our Galaxy, was proposed in ref.~\cite{ad-js} and
developed recently in ref.~\cite{ad-mk-nk}. According to the suggested model
the bulk of baryons and (almost equal) antibaryons are in the form of compact 
stellar-like objects or possible primordial black holes (PBH),
plus the observed sub-dominant practically homogeneous baryonic background,
all created by the same baryogenesis mechanism.
The amount of antimatter may be much larger than that of the
{\it known} baryons, but such ``compact'' (anti)baryonic objects 
could escape direct observations or 
the observation through their impact on BBN and CMB, 
and even make all or significant part of dark matter in the universe.  

To create compact high density baryonic and antibaryonic objects we
rely on the Affleck-Dine baryogenesis discussed above in sec.~\ref{s-bs}.
As we have mentioned, a scalar baryon $\chi$ could condense
along flat directions of its potential and accumulate a high baryonic
charge later released in the decays of $\chi$ into quarks.
However, if the window to the flat direction is open only {during 
a short period,} cosmologically small but possibly astronomically large 
bubbles with high ${ \beta}$ could be
created, occupying {a small fraction of the cosmological volume,} 
while the rest of the universe would have the normal 
baryon asymmetry (\ref{beta-obs}).

When the window to the flat direction is open, the system could undergo
a 2nd order phase transition. When the window is closed, the phase
transition should be first order. Since by assumption the window is open for
a short finite time, we call this phase transition 3/2 order.

To realize such rather unusual behavior, a very simple modification of the
potential of the Affleck-Dine field $\chi$ is sufficient.
We assume that $\chi$ has the usual Coleman-Weinberg potential~\cite{cw-pot}
plus an additional general renormalizable coupling to inflaton field $ \Phi$:
\be 
U(\chi,\Phi)  = {\lambda_1|\chi|^2 (\Phi -\Phi_1)^2} + 
\lambda_2 |\chi|^4 \,\ln (|\chi|^2/\sigma^2) +
m_0^2 |\chi^2| + (m^2_1 \chi^2 + h.c.),
\label{U-of-chi}
\ee
where $\Phi_1$ is some constant value which $\Phi$ passes in the course of
inflation but not too far from the end of inflation.
The mass parameter $ m_1$ may be complex but CP would be still conserved, 
because one can ``phase rotate'' $ \chi$ to eliminate complex parameters 
in the Lagrangian. It is essential that the last term is not invariant
with respect to $U(1)$ transformation, $\chi \rightarrow \exp(i\Theta) \,\chi$, 
and thus it breaks B-conservation.
Potential (\ref{U-of-chi}) has one minimum at $\chi =0$ for large and 
small $\Phi$ and has a deeper minimum at non-zero $\chi$ when $\Phi$ 
is close to $\Phi_1$. At that time the chances for $\chi$ to reach a high
value at the other minimum are non-negligible.

There is a simple mechanical analogy which allows to visualize the solution
of the equation of motion and the evolution of the baryonic
charge density of $\chi$. For homogeneous field $\chi = \chi(t)$
its equation of motion is the equation of the Newtonian mechanics of a 
point-like particle in potential (\ref{U-of-chi}) with the liquid friction term
${J_t^{(B)} = i\chi^\dagger \partial_t \chi + h.c. }$,
is in this language the angular momentum of the particle, ${B = J_t^{(B)}}$.
Hence we can see that during the initial period 
of inflation when the Hubble parameter was large in comparison with the effective
mass of $\chi$, $H_I^2 > |m_\chi|^2$, the amplitude of the field was near zero.
When $\Phi \approx \Phi_1$, the quantum fluctuations of $\chi$ should 
rise~\cite{fluct} and $\chi$ might reach another minimum and remained there
till the minimum disappeared at sufficiently small $\Phi$.

The probability for $\chi$ to reach the deeper minimum is determined by the
quantum diffusion at inflationary stage. It is governed 
by the diffusion equation~\cite{starobinsky}:
\be 
\frac{\partial{\cal P}}{\partial t}=
\frac{H^3}{8\pi^2}\sum_{k=1,2}\frac{\partial^2{\cal P}}{\partial\chi_k^2}+
\frac{1}{3H}\sum_{k=1,2}\frac{\partial}{\partial\chi_k}
\left[{\cal P}\frac{\partial U} {\partial\chi_k}\right]
\label{dP-dt}
\ee
where ${\chi= \chi_1+i\chi_2}$.
The inflation may be not exact and $H_I$ may depend upon time but
this does not significantly influence the spectrum of the produced bubbles with
high baryonic density.

Field $\chi$ can quantum fluctuate noticeably away from the origin when
$\Phi$ is close to $\Phi_1$ at moment $t=t_1$ 
and the effective mass of $\chi$ behave as 
{${m_{eff}^2 \approx m_0^2 + m_1^4 (t-t_1)^2}$}. Correspondingly
the dispersion is:
\be
\langle \chi^2\rangle \sim \left[ m_0^2 + m_1^4 (t-t_1)^2 \right]^{-1}.
\label{chi2-avrgd-2}
\ee

It can be shown the bubble distributions over length and mass
have the log-normal form:
\be
\frac{dn}{dM} = C_M \exp{[-\gamma \ln^2 (M/M_0)]}
\label{dn-dM}
\ee
where ${C_M}$, ${\gamma}$, and ${M_0}$ are constant parameters.

Field $\chi$ would keep its large amplitude till the second minimum remains,
even if it becomes higher than the minimum at $\chi =0$. When $\Phi$ becomes
sufficiently small and the second minimum disappears, $\chi$ would evolve
down to the minimum at $\chi = 0$. Due to an asymmetry of $U(\chi)$ with
respect to rotation in the complex $\chi$ plane, $\chi$ would start to ``rotate''
acquiring high baryonic number density.
Later the ``rotation'' of $ \chi$ would be transformed into baryonic 
number of quarks by {B-conserving decays of  $ \chi$.}

The magnitude of the baryon asymmetry inside the 
B-balls, and their size are stochastic 
quantities. {The initial phase of $\chi$ is uniform in the
interval ${[0,2\pi]}$, due to the large Hubble driving force, ${ H\gg m}$.}
The size of B-ball is determined by the remaining inflationary time after
the inflaton passed the value $\Phi_1$. The magnitude of the cosmological
baryon asymmetry ${\beta}$ could be large, especially if 
$ \chi$ decayed much after the inflaton decay and the entropy dilution was absent.

In the simplest version of the model both positive and negative values of 
${ \beta}$ in the baryon rich bubbles are equally probable. 
The background uniform baryon asymmetry with 
{${\beta = 6\cdot 10^{-10}}$} in the main part of the universe may be created by 
the same field $\chi$ which did not penetrate to the second minimum of the
potential but ``lived'' near zero. To this end an explicit CP-violation in the  
$\chi$ sector is necessary. Another possibility of creation of the small and
homogeneous baryon asymmetry in the bulk of the universe
is one of the mechanisms enumerated in sec.~\ref{s-bs}. 

\section{Baryonic inhomogeneities and their evolution \label{s-inhom}}

According to the scenario described above,
the universe looks as a huge piece of swiss cheese or better to
say, as ``anti'' swiss cheese because in the bulk with the normal baryon number 
density there are small dense bubbles with much larger baryon 
or antibaryon number density,  with {${|\beta| \sim 1}$},
but not empty holes as in the cheese. The mass of those high B objects can be
of the order of stellar mass or even larger or much smaller, as we see below.
Despite their small size
the mass fraction of the bubbles could be comparable or even larger than
the observed baryonic mass fraction. 

Initially the density contrast between the bubbles with high values of $\chi$
and the bulk with $\chi \sim 0$ was small, if the energy density of $\chi$ was
much smaller than the energy density of the inflaton. This density contrast 
remained constant while the matter inside and outside the B-bubbles were
relativistic. Later when the mass of $\chi$ came into play, the matter inside
the bubbles with a large amplitude of $\chi$ became nonrelativistic and the
density contrast started to rise. The rise continued till $\chi$ decayed into
light quarks and/or leptons and the matter inside became relativistic as in the
bulk of the universe.

The second period of the rising perturbations took place after
the QCD phase transition at $T=T_{QCD} \sim 100$ MeV,
when relativistic quarks confined to make nonrelativistic nucleons.

If ${\delta\rho /\rho = 1}$ at horizon crossing, primordial black holes 
(PBH) would be formed. The mass inside horizon at cosmological time $t$
is equal to:
\be
M_{hor} \approx  m_{Pl}^2 t\approx 10^{38} \rm{ g}\, ({ t}/\rm{ sec})
\approx 10^{5} M_\odot (t/sec),
\label{M-hor}
\ee
where $M_\odot$ is the Solar mass.
Time is related to the temperature by a simple approximate equation
$t/{\rm sec} = \left( T/ {\rm MeV}\right)^{-2}$. 
Hence for ${T=10^{8}}$ GeV the PBH mass would be  ${10^{16} }$ g.
{Perturbations with ${\delta \rho/\rho < 1}$ might still
make PBH due to subsequent matter accretion.}
If PBH had not been formed, the subsequent evolution of the B-bubble depends 
upon the relation between their mass and the Jeans mass, see below.

At the moment of QCD phase transition the mass inside horizon is about $M_\odot$, 
while during big bang nucleosynthesis (BBN) the mass inside horizon varies from
$10^5 M_\odot$ to to ${10^{7} M_\odot}$. One should keep in mind that
compact objects (not BH) with smaller masses could be formed too.

Initial inhomogeneous $\chi$ and/or
$ \beta$ led to large isocurvature perturbations. The
amplitude of such perturbations is strongly restricted by BBN and
by CMBR at about 10\% level, but these bounds are valid 
at much larger scales, bigger than galactic size.

The amplitude of relative density perturbations, 
when they entered horizon after the QCD phase transition is equal to:
\be 
r_B=\frac{\delta\rho}{\rho} = \frac{\beta n_\gamma m_p}{(\pi^2/30) g_* T^4}
\approx 0.07 \beta\,\frac{ m_p}{T}. 
\label{r-B}
\ee
For $\beta = 1$ the density contrast at horizon crossing would be of the
order of unity at $T \approx 70$ MeV, and $M_{hor} \approx 10^2 M_\odot$. 
In this way an early formation of very heavy PBH, up to superheavy ones observed
in all large galaxies can be understood. At the present time the mechanism  
of their creation is unknown, for a review see ref.~\cite{doku}

Thus the bubbles with high baryonic number density could naturally form
PBH with masses either $10^{16}$ g, or somewhere in this region, or solar
mass, or much heavier PBH. Of course BH made of matter or antimatter are 
indistinguishable because baryonic charge does not create any long range 
forces, for a review see~\cite{ad-long}. However, 
anti-BH may be surrounded by anti-atmosphere if $ \beta$ {slowly} 
decreases. This makes them potentially observable antimatter objects.

If $ {M_0 \sim M_\odot}$, eq. (\ref{dn-dM}), some of the high $ \beta$ bubbles 
might form stellar type objects in the early universe, more or less at the 
era of BBN. Most probably these stars are now evolved and dead or have low 
luminosity. Both such stars and PBHs may make a considerable contribution
into the cosmological dark matter.

Nonrelativistic
baryonic matter started to dominate inside the bubbles at 
\be 
T =T_{in} \approx 65 \, \beta \,{\rm MeV}
\label{T-beta}
\label{T-eq}
\ee
The mass inside a baryon-rich bubble with radius $R_B$
at the radiation dominated stage was 
\be 
M_B \approx
2\cdot 10^5 \, M_\odot (1+r_B) \left(\frac{R_B}{2t}\right)^3\,
\left(\frac{t}{\rm sec}\right) 
\label{M-B}
\ee
The mass density at the onset of matter dominated (MD) stage was:
\be
\rho_B 
\approx 10^{13} \beta^4 \,\, {\rm g/cm}^3\, .
\label{rho-B}
\ee
The bubbles with ${{ {\delta\rho}/{\rho}<1}}$ but with
$${{M_B>M_{Jeans}}} $$
at horizon {would decouple from the cosmological expansion} and
form compact stellar type objects or lower density clouds which
could survive against early annihilation.

For example, for a solar mass bubble the mass density is
\be 
\rho_B = \rho_B^{(in)} (a_{in}/a)^{3} \approx 6\cdot 10^5\,\,
{\rm g/cm^3}
\label{rho-Bb}
\ee
and the radius is $ {R_B \approx 10^9}$ cm. The
temperature when ${M_J=M_\odot}$ is: 
\be 
T\approx T_{in} (a_{in}/a)^2 \approx 0.025\,\,{\rm  MeV}. 
\label{T-msun}
\ee
This bubble is similar to the red giant core and its evolution
should be also similar, but with an important difference that initially
the external pressure was larger then the internal one.

There are three sources of energy release by such objects at an early stage:\\
{1. Cooling down because of high internal temperature, ${T\sim 25}$ keV.}\\
{2. Annihilation of surrounding matter on the surface.}\\
{3. Nuclear reactions inside.} 

1. The cooling time is determined by the  photon diffusion time:
\be 
t_{diff} 
\approx 2\cdot 10^{11}\,{\rm sec} 
\left(\frac{M_B}{M_\odot}\right)\,\left(\frac{\rm sec}{R_B}\right) 
\left(\frac{\sigma_{e\gamma}}{\sigma_{Th}}\right)
\label{t-diff}
\ee
Correspondingly the luminosity is {${L\approx 10^{39}}$ erg/sec.} 
With the thermal energy stored inside such B-ball equal to
\be 
E_{therm}^{(tot)} = 3T M_B / m_N 
\approx  1.5\cdot 10^{50} {\rm erg}
\label{E-therm}
\ee
the life-time with respect to the cooling would be about $10^{11}$ s.
In the extreme case of all DM made of such B-balls i.e. for
${ \Omega_{BB} = 0.25}$, the thermal keV photons would make
{${10^{-4}-10^{-5}}$ of CMBR,} {red-shifted today to the
background light.} 

{2. Nuclear helium burning, } (similar to a red giant core): 
$ {3He^4 \rar C^{12}}$, however with larger T by the factor
$ {\sim 2.5}$. Since the luminosity  strongly depends upon the core
temperature, $ {L\sim T^{40}}$, the life-time with respect to this process
would be very short. The total energy influx would be below
${10^{-4}}$ of CMBR if {${\tau < 10^9}$ s.} However, such an intensive
nuclear burning could lead to the B-ball explosion and creation of a
solar mass anti-cloud.

{3. Annihilation on the surface.}
(Anti)proton mean free path before 
recombination is small:
\be
l_p = {1}/{(\sigma n)} \sim {m_p^2}/({\alpha^2\,T^3}) = 0.1\, cm\, 
\left({MeV}/{T}\right)^3
\label{l-p1}
\ee
After recombination the number of annihilation
on one B-ball per unit time would be: 
\be 
\dot N = 
10^{31} V_p \left({T}/{ 0.1\,\,{ {eV}}}\right)^3
\left({R_B}/{10^9\,\,{{ cm}}}\right)^2,
\label{dot-N-recomb}
\ee
With maximum allowed number of B-balls it could create the energy density 
of 100 MeV photons, properly red-shifted by
today, not more than ${10^{-15}}$ of CMBR.

Thus we see that compact anti-objects could survive in the early universe,
even if they are not PBHs.
A kind of early dense stars might be formed with 
initial pressure outside larger than that inside.
Such very first stars might evolve quickly and, in
particular, make early SNs,  enrich the universe with heavy
(anti)nuclei and re-ionize the universe.
The energy release from stellar like objects in the early
universe is small compared to CMBR.
Such objects are not dangerous for BBN since they occupy a very small fraction
of the total cosmological volume.

\section{Observational effects of antimatter in the Galaxy \label{s-antigal} }

Here we will discuss possible effects induced by antimatter objects in 
our neighborhood. The presentation is based on our paper~\cite{cb-ad}. For
other works on similar issue see refs.~\cite{fargion}. We would not
dwell on a particular theoretical model but still keep in mind the possibility
of compact antimatter objects discussed in the previous sections. The list of
possible astronomically significant antimatter objects, which may live
in the Galaxy includes:\\
{1. Gas clouds of antimatter.}\\
{2. Isolated anti-stars, maybe already dead}\\
{3. Anti stellar clusters.}\\
4. Anti black holes with possible anti-atmosphere around.\\
These anti-objects may be inside galaxies or outside them. They can be concentrated
in the galactic halos or be in the intergalactic space.
All the options are open

The observational signatures of such objects are more or less trivial. They could be
100 MeV gamma background or compact sources of such gamma radiation from 
$\bar p p$ annihilation. There could be excessive antiprotons or positrons
in the cosmic rays. 
The 0.511 MeV line from $e^+e^-$--annihilation may also be
a signature of abundant cosmic antimatter. Probably the strongest indication for
cosmic antimatter would be an observation of anti-nuclei starting from anti-$^4 He$
to heavier ones. Among more difficult to observe effects are the photon polarization
from the synchrotron radiation and the fraction of neutrino versus antineutrino
from supernova explosion.

The antimatter objects in the galaxy are floating in the galactic gas of protons
which should annihilate with anti-protons in such objects. This would give rise
to gamma-radiation from ${{ \bar p p \rar pions}}$ and 
${{ \pi^0 \rar 2\gamma}}$ (${{ E_\pi \sim 300}}$ MeV) and from 
${ e^+e^-}$-annihilation originating from ${{ \pi^\pm}}$-decays
and also 0.511 MeV photons from the "original" positrons in the B-ball. 

The astronomically large antimatter objects can be separated into two classes:
gas clouds and compact stellar-like objects.
We define the gas clouds of antimatter as objects for which the
mean free path of protons, ${{ l_p}}$, is larger than 
the size of the (anti)cloud, ${{ l_c\equiv l_B}}$.
\be
l_p = {1}/({\sigma_{tot} n_{\bar p}}) = 10^{24}\, {\rm cm} \,
\left({cm^{-3}}/{n_{\bar p}}\right)\, 
\left({barn}/{\sigma_{tot}}\right), 
\label{l-p2}
\ee
Here $n_{\bar p}$ is the number density of antiprotons in the cloud.
According to the model discussed above 
it is natural to expect that $n_{\bar p} \gg n_p \approx 1/$cm$^3$, where
$n_p$ is the average number density of protons in the Galaxy.

If the relative velocity, $v$, of the $\bar p p$ is small in comparison
with the fine structure constant, $\alpha = 1/137$, the cross-section
of the annihilation is amplified by the Sommerfeld-Sakharov~\cite{som-sakh}
factor:
\be
\sigma \rar \sigma \,\frac{2\pi\alpha/v}{1- \exp (-2\pi \alpha/v)}
\label{som-sakh}
\ee

Galactic protons would penetrate into an anti-cloud and annihilate inside
its whole volume. It leads to a high efficiency of the annihilation. The number of
annihilations per unit time is 
\be
\dot n_p = v\sigma_{ann} n_p n_{\bar p}
\label{dot-n-p}
\ee
Correspondingly the total number of annihilations is equal to 
{${ \dot N_p = 4\pi l_c^3\,\dot n_p /3}$.}
The total number of ${ \bar p}$ in the cloud is
{${{ N_{\bar p} = 4\pi l_c^3 n_{\bar p}/3}}$.}
So a low density or small clouds would not survive in a galaxy, but they could 
survive in the halo. The life-time of an antimatter cloud submerged into 
a sea of protons with density $n_p$ and velocity $v_p$ is: 
\be
{ \tau =  10^{15} \,\, sec\,\, 
\left(\frac{10^{-15}cm^3/s} {\sigma_{ann} v}\right)\,
\left(\frac{cm^{-3}}{n_p}\right),}
\label{tau-c}
\ee
if supply of galactic protons is sufficient.
Indeed, the proton flux into an anti-cloud:
\be
F = 4\pi l_c^2 n_p v = 10^{35}\,sec^{-1}\left({n_p}/{cm^3}\right)
\left({l_c}/{pc}\right)^2
\label{F}
\ee
This flux is sufficient to destroy the anti-cloud in ${ 10^{17}}$ sec if:
\be
\left({n_{\bar p}/}{cm^3}\right)
\left({l_c}/{pc}\right) < 3\cdot 10^4
\label{frac-}
\ee
The luminosity for volume annihilation is quite high:
\be
L_\gamma^{(vol)} 
\approx
 10^{35}\,{\rm \frac{erg}{s}}\, 
 \left(\frac{R_B}{0.1\,{\rm pc}}\right)^3 
 {{\left( \frac{n_p}{{\rm 10^{-4}\,cm}^{-3}}\right)}}
 {{ \left(\frac{n_{\bar p}}{10^4 {\rm cm}^{-3}}\right). 
}}\label{L-gamma-vol2}
\ee
It would create the flux of 100 MeV photons 
on the Earth at e.g. distance of one kpc 
{${10^{-5}\gamma/{\rm s/cm}^2 }$} or 
{${10^{-3}{\rm Mev/\,s/cm}^2 }$ },
similar to the observed cosmic background, 
{${10^{-3}/{\rm MeV/s/cm}^2 }$.}

If the density of antimatter inside a B-rich bubble is so high that the
proton mean free path is smaller than the bubble size,  $ {l_{free} < l_B}$,
the annihilation would proceed in a narrow shell near the surface.
All that hits the surface annihilate, but the ``surface-to-volume'' ratio is
very small and the annihilation is not efficient enough to destroy the object.
The total luminosity with respect to the surface annihilation is: 
\be
L_{tot}  = 2m_p \cdot 4\pi\,R_B^2\,n_p v 
\approx 10^{27}\,{\rm \frac{erg}{sec}}\,
\left(\frac{n_p}{cm^3}\right)\left(\frac{R_B}{l_\odot}\right)^2 
\label{L-tot}
\ee
The fraction of the annihilation products into gamma-rays is about 
20-30\%. 

Much more energetic may be the annihilation of galactic protons with
antiprotons from the stellar wind produced by an anti-star. The mass loss
by an anti-star can be parametrized as:
\be{{
\dot M = 10^{12} W\,{\rm g/sec}
}}\label{dot-M}
\ee
where ${ W=\dot M /\dot M_{\odot}}$ is the mass loss of an anti-star in units of 
that for the Sun. If all ``windy'' particles annihilate, the luminosity
per anti-star would be: 
\be
{{{ L= 10^{33} W \,\, {\rm erg/sec}. }}}
\label{L-wind}
\ee
These compact objects would be quite bright sources of energetic, $\sim 100$ MeV,
photon radiation. 
The observational restrictions permit to impose the limit on the
number density of anti-stars to that of the usual solar type stars:
\be
N_{\bar S} / N_S \leq 10^{-6} W^{-1},
\label{N-bar}
\ee
from the total galactic luminosity in 100 MeV photons, 
${ L_\gamma = 10^{39} erg/s}$, and from the flux of the positron 
annihilation line ${ F \sim 3\cdot 10^{-3}/cm^2/s}$.
It is natural to expect that $ {W\ll 1}$ because the primordial anti-stars 
are, most probably, already evolved.

The stellar wind would populate galaxy with anti-nuclei. Their number density  
is bounded from above by the density of
``unexplained'' ${ \bar p}$ and the fraction of anti-nuclei in
stellar wind with respect to antiprotons.
It may be the same as in
the Sun, but if anti-stars are old and evolved, this number must be much 
smaller. Heavy anti-nuclei from primordial
anti-SN explosion may be abundant but their
ratio to ${ \bar p}$ can hardly exceed the same for normal SN.
Explosion of anti-SN would create a large cloud of antimatter, which
should quickly annihilate producing vast energy - a spectacular event.
However, most probably such stars are already dead and anti-SN might
explode only in very early galaxies or even before them.
The observational bounds on the antihelium-helium ratio~\cite{anti-he}
impose:
\be
N_{\bar S} / N_S  = (\bar{He}/He) \leq 10^{-6},
\label{N-He}
\ee
if anti-stars are similar to usual stars, though most probably 
they are not and the limit is weaker.

As we have already mentioned, the abundances of light elements created 
during BBN in baryon-rich bubbles should be significantly different from
those produced in the bulk with normal $\beta$, eq.~(\ref{beta-obs}).  
If ${\beta  \gg 10^{-9} }$, the light (anti)element 
abundances would be anomalous: much less 
anti-deuterium and more anti-helium should be produced. For the
calculations of light and heavier element abundances in the bubbles with 
anomalously high $\beta$ see ref.~\cite{matsuura}.
This opens a possibility for search of antimatter objects. If there are
some regions in the sky with anomalous chemistry, they have a good chance to be 
made of antimatter. Of course, according to the discussed above mechanism of 
antimatter creation, there is 50\% probability that such regions consist of
normal matter with anomalously high ratio ${ n_B /n_\gamma}$.
Still if such a cloud or compact object is found, the search for
annihilation there has non-negligible chance to be successful.

Compact objects made of antimatter could be efficient sources of
cosmic positrons. Indeed, the gravitational proton capture by an anti-star 
is more efficient than the capture of electrons due to smaller
mobility of the latter in the interstellar medium, see
e.g.~\cite{cb-ad-ap}. The anti-star would be neutralized by a forced positron 
ejection. The process would be most efficient in galactic center where the 
proton number density, ${n_p}$ is large. The 0.511 MeV positron annihilation
line must be accompanied by wide spectrum ${\sim 100}$ MeV radiation coming from
$\bar p $--annihilation.

There could be also quite spectacular but rare event of star-antistar
collisions. The collision of similar mass star-antistar would resemble
$ \gamma$-bursters. The energy release at this collision is
\be 
\Delta E \sim 10^{48}\,{\rm erg}\, \left(\frac{M}{M_\odot}\right)
\left(\frac{v}{10^{-3}}\right)^2,
\label{Delta-E}
\ee
where $v$ is the relative velocity of the colliding stars and $M$ is their
mass. The annihilation pressure would push the stars apart with the
collision time being about 1 second.
The radiation should be emitted in the narrow disk on the boundary of the
colliding stars but not in jets.

The collision of a compact anti-star with a red giant would lead to penetration
of the anti-star into the giant. The compact anti-star would travel inside 
creating an additional energy source. As a result the 
color and luminosity of the red giant would change. The characteristic 
time for  such a process is about one month. The additional energy 
release during this period is ${\Delta E_{tot} \sim 10^{38}}$ erg.
The presented estimates are very rude and more detailed calculations are in order.

Another two body violent phenomenon could be generated by the
transfer of matter in a binary system of a star and 
an anti-star. It would lead
to the effects similar to hyper-nova explosion with hard spectrum of emitted
photons.

Among more subtle effects is the photon polarization from synchrotron
radiation. Since positrons are predominantly 
``right handed'', the same is transferred to the bremsstrahlung. 
Another potentially observable one is registration of neutrinos from
supenova explosion: the first burst ${\nu}$ from SN consists of neutrinos
while that from anti-supernova consists of antineutrinos.

The bubbles with high baryonic (and anti-baryonic)
number density should contribute into 
cosmological dark matter. If their mass spectrum is centered near relatively
small mass, say, from $10^{15}$ g to planet masses they would not emit 
light and would behave as normal cold dark matter. Some or many of them could
form PBHs. They might even dominate in DM. For the observational bounds on
PBH see e.g. ref.~\cite{carr}.
Heavier B-bubbles open much more exciting possibilities. They could make
very first stars, which would enrich interstellar space with anti-elements.
They may manifest themselves as MACHOs~\cite{macho} and make a considerable
contribution into dark matter. The bounds on the possible fraction of dark 
matter in the form of stellar (or similar) mass objects can be found
in refs.~\cite{macho,bh-dm}.

\section{Conclusion \label{s-concl}}

We have shown that there exists a realistic possibility that the
amount of antimatter in the universe may be noticeably larger than the 
amount of the observed matter. Moreover, the antimatter objects can be not
far away near horizon but may populate our Galaxy, as well as all other ones.
They still escaped observations, because they are mostly compact, 
but with new more sensitive instruments the 
odds for their discovery are non-negligible.

As a by-product the suggested scenario presents a 
feasible mechanism of creation of superheavy BH in the galactic centers
which might be seeds for galaxy formation. To the best of my knowledge
this is the only mechanism  of early quasar formation with evolved chemistry,
which is one of the mysteries of the standard cosmology.

As cold DM particles B-bubbles should be abundant in the galactic haloes.
Since no shining stars are observed in the halo, it means that 
the high B compact objects are mostly dead or low luminosity stars. The stellar
wind from them is low or absent. However, annihilation of background protons on  
their surface should exist and compact sources of 100 MeV radiation may be 
observed. Despite of low intensity of the stellar wind, the B-bubbles 
could eject anti-nuclei into interstellar space during early period of their
evolution.
Not only ${ ^4 \bar{He}}$ is worth to look for but 
also heavier anti-elements. Their relative abundances should be similar 
to those observed in SN explosions.
Regions with an anomalous abundances of light elements are suspicious
that they are dominated by antimatter and there may be anti-elements.
These regions may be sources of gamma radiation from $\bar p p$ and
$e^+e^-$--annihilation. 

Discovery of cosmic antimatter looks as the unique chance to
establish what baryogenesis scenario was realized in the universe. All
the usual scenarios deal with only one number, $\beta$, and the measurement
of this one number does not allow to distinguish between different mechanisms.

Possibly 0.511 MeV $e^+e^-$--annihilation line line from the
galactic center and maybe even from the galactic halo~\cite{bulge,halo}
are the first positive indications to cosmological antimatter.


\end{document}